\documentclass[letterpaper,conference]{IEEEtran}

\usepackage[T1]{fontenc}
\usepackage[utf8]{inputenc}
\usepackage{cite}
\usepackage{amsmath,amssymb,esint}
\usepackage{graphicx,xcolor}
\usepackage[super]{nth}
\usepackage{psfrag}
\usepackage{pbox,ragged2e}
\usepackage{balance,multicol}
\usepackage{placeins}

\usepackage[caption=false,font=footnotesize]{subfig}
\usepackage{textcomp}
\usepackage{gensymb}
\usepackage{dsfont}
\usepackage{bbm}
\usepackage{upgreek} 

\usepackage{tikz,pgf,pgfplots}
\pgfplotsset{compat=1.14}

\DeclareMathOperator*{\argmin}{arg\,min}

\newcommand{\f}[2]{\frac{#1}{#2}} 

\newcommand{\fp}[2]{\f{\partial #1}{\partial #2}} 

\newcommand{\diag}{\mathrm{diag}} 

\newcommand{\eye}{\mathbf{I}}

\newcommand{\mtx}[2]{\left[\begin{array}{#1}#2\end{array}\right]}

\newcommand{\Tr}{^\mathrm{T}}

\newcommand{\re}{\mathrm{Re}}
\newcommand{\im}{\mathrm{Im}}	

\newcommand{\p}{\mathbf{p}}
\newcommand{\oFIX}{\mathbf{o}}
\newcommand{\ag}{_\mathrm{ag}}
\newcommand{\an}{_\mathrm{an}}

\newcommand{\fieldLOS}[1]{{\bf v}_{\text{dir}#1}}
\newcommand{\fieldNLOS}[1]{{\bf v}_{\text{mp}#1}}

\newcommand{\model}{^\text{model}}
\newcommand{\meas}{^\text{meas}}

\newcommand{\estParam}{{\boldsymbol\psi}}
\newcommand{\FIM}{\pmb{\mathcal{I}}}

\newcommand{\um}{\,\text{\textmu}\mathrm{m}}

\newcommand{\e}{\mathbf{e}}

\newcommand{\coeff}{\alpha}

\newcommand{\nf}{_\text{NF}}
\newcommand{\ff}{_\text{FF}}

\newcommand{\bnf}{\boldsymbol\upbeta\nf}
\newcommand{\bff}{\boldsymbol\upbeta\ff}

\newcommand{\unit}[1]{\,\mathrm{#1}}

\newcommand\ModelError{\epsilon}
\newcommand\ModelErrorVec{{\boldsymbol\ModelError}}

\title{Practical Accuracy Limits of Radiation-Aware Magneto-Inductive 3D Localization}
\author{%
\IEEEauthorblockN{Gregor Dumphart, Henry Schulten, Bharat Bhatia, Christoph Sulser, and Armin Wittneben}
\IEEEauthorblockA{Communication Technology Laboratory, ETH Zurich, Switzerland\\
Email: \{dumphart, schulten\}@nari.ee.ethz.ch, bhatiab@ethz.ch, \{sulser, wittneben\}@nari.ee.ethz.ch, }}

\begin{document}

\maketitle

\begin{abstract}
The key motivation for the low-frequency magnetic localization approach is that magnetic near-fields are well predictable by a free-space model, which should enable accurate localization. Yet, limited accuracy has been reported for practical systems and it is unclear whether the inaccuracies are caused by field distortion due to nearby conductors, unconsidered radiative propagation, or measurement noise. Hence, we investigate the practical performance limits by means of a calibrated magneto-inductive system which localizes an active single-coil agent with arbitrary orientation, using 4\,mW transmit power at 500\,kHz. The system uses eight single-coil anchors around a 3m\,$\bf\times$\,3m area in an office room. We base the location estimation on a complex baseband model which comprises both reactive and radiative propagation. The link coefficients, which serve as input data for location estimation, are measured with a multiport network analyzer while the agent is moved with a positioner device. This establishes a reliable ground truth for calibration and evaluation. The system achieves a median position error of 3.2\,cm and a 90th percentile of 8.3\,cm. After investigating the model error we conjecture that field distortion due to conducting building structures is the main cause of the performance bottleneck. The results are complemented with predictions on the achievable accuracy in more suitable circumstances using the Cram\'er-Rao lower bound.
\end{abstract}
\section{Introduction}\label{sec:intro}
In dense propagation environments, radio localization faces severe challenges from radio channel distortions such as line-of-sight blockage or multipath propagation \cite{BuehrerPIEEE2018,MazuelasTSP2018,SchultenVTC2019}. Its use inside of buildings, underground, or for medical applications is thus drastically limited.
Magnetic near-fields, in contrast, are hardly affected by the environment as long as no major conducting objects are nearby \cite{Barr2000,SheinkerTIM2018,PaskuTIE2016,AbrudanJSAC2015,ArumugamAPM2014,KyprisTGRS2016}. Thus, the magnetic near-field at some position relative to the source (a driven coil or a permanent magnet) can be predicted accurately with a free-space model. This enables the localization of an agent in a setup with stationary coils of known locations (i.e. anchors) \cite{PaskuTIE2016,ArumugamAPM2014,AbrudanJSAC2015,KyprisTGRS2016,DaiSENS2018,DumphartPIMRC2017}. In particular, position and orientation estimates can be obtained by fitting a dipole model to measurements of induced voltage or of a related quantity \cite{ArumugamAPM2014,AbrudanJSAC2015,KyprisTGRS2016,DaiSENS2018,DumphartPIMRC2017}.

While these circumstances present the prospect of highly accurate magneto-inductive localization, only mediocre accuracy has been reported for practical systems, with a relative position error of at least $2\%$ \cite{ArumugamAPM2014,SheinkerTIM2018,PaskuTIE2016,AbrudanJSAC2015,KyprisTGRS2016}. Possible sources of error comprise noise and interference, quantization, signal model inadequacies (e.g. weak coupling and dipole assumptions), poorly calibrated model parameters, unconsidered radiative propagation (direct path and multipath), and field distortion due to induced eddy currents in nearby conductors.
The latter can be captured with an image source model in the case of a conducting ground or reinforcing bars, which led to appreciable but limited improvements of the localization accuracy in \cite{ArumugamAPM2014,KyprisTGRS2016}. The authors of \cite{ArumugamAPM2014} note that radiative propagation should be considered by future solutions. To the best of our knowledge, it is currently unknown which error source causes the accuracy bottleneck for existing systems.

In this paper we make the following contributions. We extend the common magnetoquasistatic dipole model to a complex formulation that accommodates both reactive and radiative propagation. Moreover, we compensate contributions from radiative multipath propagation by exploiting the large wavelength and calibration. A previously proposed algorithm for position and orientation estimation \cite{DumphartPIMRC2017} is consequently adapted to this model. Based thereon, we present a system implementation for magneto-inductive 3D localization whereby anchors and agent use flat spiderweb coils tuned to $500\unit{kHz}$. We evaluate the achievable accuracy in an office setting after thorough calibration. To allow for a rigorous evaluation, all measurements are made with a multiport network analyzer, i.e. the agent is tethered and furthermore mounted on a controlled positioner device. We investigate the different sources of error and conjecture that field distortion due to reinforcement bars causes the accuracy bottleneck. We estimate the potential accuracy in more ideal circumstances using the Cram\'er-Rao lower bound (CRLB) on the position error and a Gaussianity assumption.\\[-3.5mm]

\subsubsection*{Related Work}
An average error of $11\unit{cm}$ is reported by \cite{SheinkerTIM2018} for a $(7\unit{m})^2$ setup with a triaxial magnetometer and a two-coil anchor ($4\unit{W}$ power, $270$ turns, $2071\unit{Hz}$) in a magnetic laboratory. They mention calibration and interference as possible imperfections.
Pasku et al. \cite{PaskuTIE2016} achieve $30\unit{cm}$ average error (2D) over a $15\unit{m} \times 12\unit{m}$ office space using coplanar coils ($20$ turns with $7\unit{cm}$ radius, $0.14\unit{W}$, $24.4 \unit{kHz}$).
A 50-turn coil wound around a football, driven with $0.56\unit{W}$ at $360\,\mathrm{kHz}$, is localized in a $(27\unit{m})^2$ area with $77\unit{cm}$ mean error in \cite{ArumugamAPM2014}. They use large receivers and techniques to mitigate self-interference from induced signals in the long cables.
Abrudan et al. \cite{AbrudanJSAC2015} report $30\unit{cm}$ mean accuracy in undistorted environments and $80\unit{cm}$ otherwise over about $10\unit{m}$ distance. They use triaxial 80-turn coils at $2.5\unit{kHz}$.
In our work \cite{DumphartPIMRC2017}, we improved the non-linear least squares approach to single-coil agent localization, thereby vastly improving robustness and runtime, and used Cram\'er-Rao bounds for accuracy projections in thermal noise which was considered as the only limitation.

This paper is structured as follows. Sec.~\ref{sec:model} presents the signal model and Sec.~\ref{sec:algo} states the employed position estimation algorithm. In Sec.~\ref{sec:setup} we describe the chosen coil design and system setup. We evaluate the achieved performance and the observed model error in Sec.~\ref{sec:eval} and compare them to the CRLB-based accuracy projections in Sec.~\ref{sec:bounds}. Then Sec.~\ref{sec:summary} concludes the paper.

\section{Signal Model versus Measurement}\label{sec:model}
\begin{figure}[!ht]
\centering
\includegraphics[width=\columnwidth]{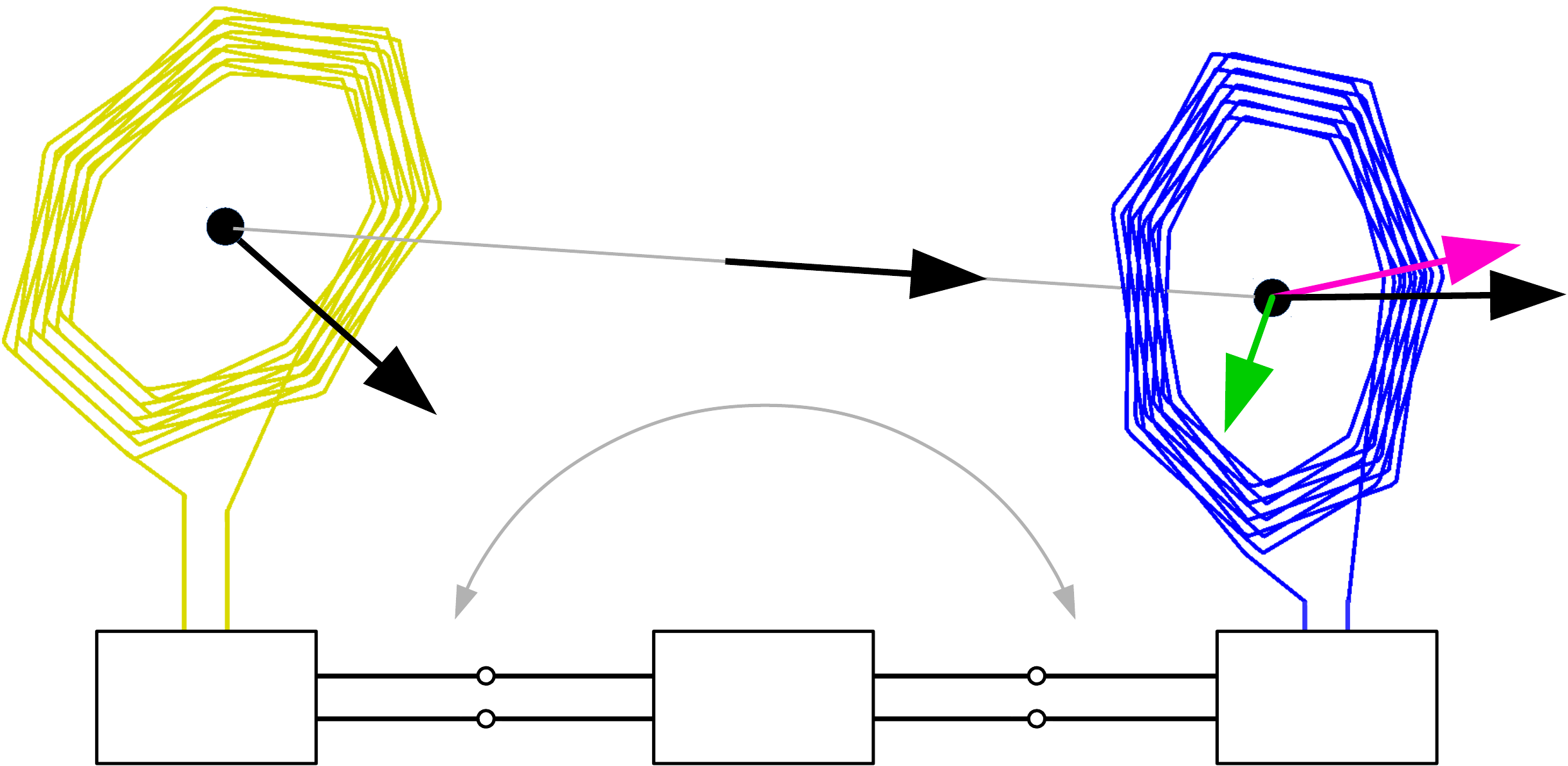}
\put(-163,88.5){$d$}
\put(-118.5,85.5){$\e$}
\put(-143,46){$h \in \mathbb{C}$}
\put(-161.5,32){$\hat{=}$\small{ $S_{21}$-Parameter}}
\put(-193,-3){\small{$50\,\Omega$ Port}}
\put(-104,-3){\small{$50\,\Omega$ Port}}
\put(-222,96){$\p\an$}
\put(-182,68){$\oFIX\an$}
\put(-54,85){$\p\ag$}
\put(-18,65){$\oFIX\ag$}
\put(-50.5,54.5){\color[rgb]{0,0.8,0}$\bff$}
\put(-21,92){\color[rgb]{1,0,0.8}$\bnf$}
\put(-141.5,12.0){\scriptsize Network}
\put(-141.9,5.0){\scriptsize Analyzer}
\put(-52.0,12.0){\scriptsize Matching}
\put(-51.0,5.0){\scriptsize Network}
\put(-232.0,12.0){\scriptsize Matching}
\put(-231.0,5.0){\scriptsize Network}

\caption{Link between the agent and an anchor with spiderweb coil geometries and description of the geometrical quantities. The employed methodology for tethered measurement of the channel coefficient $h \in \mathbb{C}$ is illustrated.}
\label{fig:LinkGeometry}
\end{figure}

We consider the link between the agent coil with center position $\p\ag \in \mathbb{R}^3$ with axis orientation $\oFIX\ag \in \mathbb{R}^3$ (unit vector) and an anchor coil with center position $\p\an \in \mathbb{R}^3$ and axis orientation $\oFIX\an \in \mathbb{R}^3$ (unit vector). For ease of exposition we discard the anchor index. The link geometry is depicted in Fig.~\ref{fig:LinkGeometry}. Both coils are conjugately matched at operation frequency $f_\mathrm{c}$ and thus resonant.
We consider the channel coefficient $h \in \mathbb{C}$ at $f_\mathrm{c}$ and describe it with the model
\begin{align}
h\model &= \coeff ( \fieldLOS{} + \fieldNLOS{} )\Tr \hspace{.2mm} \oFIX\ag 
\label{eq:SignalModelGlobal}
\end{align}
whereby all occurring quantities are unitless ($h\model$ applies to power waves). To understand \eqref{eq:SignalModelGlobal} intuitively it is helpful to think of a transmitting anchor, although the model holds for either link direction. The field vectors $\fieldLOS{}, \fieldNLOS{} \in \mathbb{C}^3$ describe direct-path propagation and multipath propagation, respectively. The former is given by%
\footnote{The formula follows from the fields of a circular loop antenna in free space \cite[Eq.~5-18]{Balanis2016}, a reformulation of the trigonometric expressions with projective geometry, and assuming a spatially constant field across the receive coil, cf. \cite[Eq.~12]{DumphartWCNC2019}. It also holds for infinitesimal dipoles on both ends \cite{Balanis2016}.}
\begin{align}
\fieldLOS{} = j e^{-jkd} \! \left( \left( \f{1}{(kd)^3} + \f{j}{(kd)^2} \right) \bnf + \f{1}{2kd}\,\bff \right)
\label{eq:DirectPath}
\end{align}
whereby the near-field and far-field geometries are given by
\begin{align}
& \bnf = \f{1}{2}\!\left( 3\e\e\Tr - \eye_3 \right) \oFIX\an , \\
& \bff = \left( \eye_3 - \e\e\Tr \right) \oFIX\an \, .
\end{align}
We use wave number $k = 2\pi f/c$, distance $d = \|\p\ag - \p\an\|$, and direction vector $\e = (\p\ag - \p\an)/d$  from anchor to agent.
The technical parameters are subsumed in
\begin{align}
\coeff = \f{\mu\hspace{.2mm}S\ag S\an \nu\ag \nu\an}{\sqrt{4 R\ag R\an}} k^3 f_\mathrm{c} \cdot \xi
\label{eq:coeff}
\end{align}
with permeability $\mu$, coil surface areas $S\ag$ and $S\an$ (the mean over all turns), coil turn numbers $\nu\ag$ and $\nu\an$, and coil resistances $R\ag$ and $R\an$.

\begin{figure}[!ht]
  \centering
  \includegraphics[width=\columnwidth,trim=0 0 0 0,clip=true]{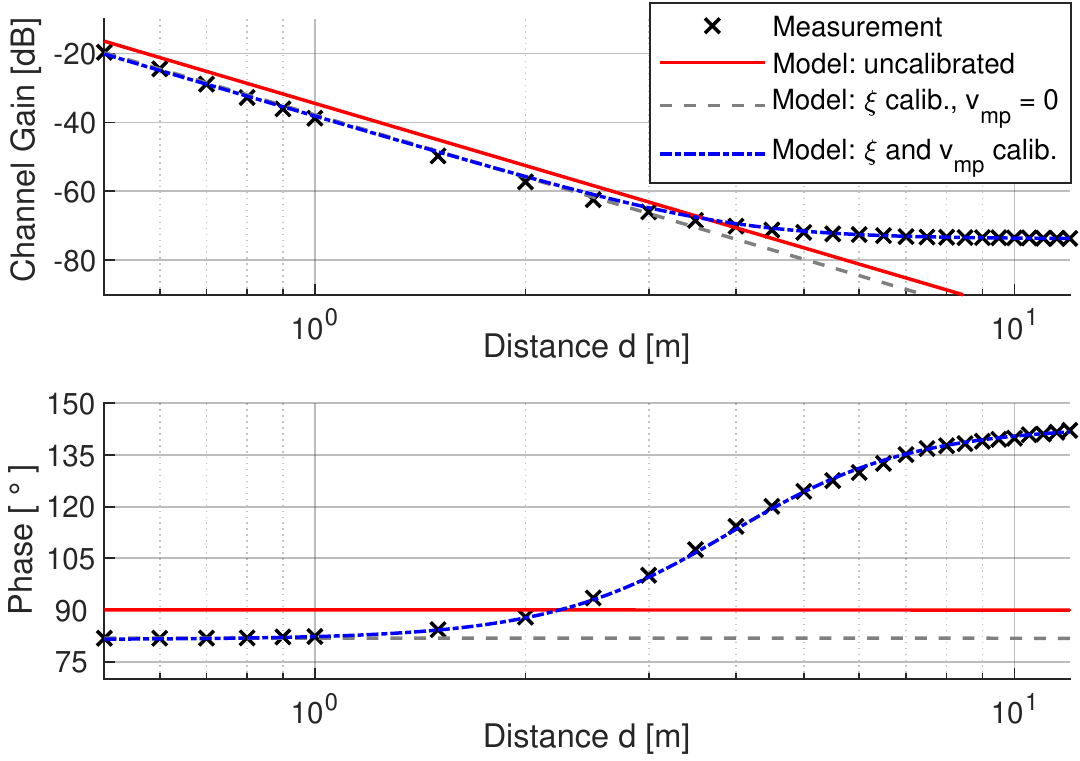}
  \caption{Channel coefficient $h \in \mathbb{C}$ of a coaxial link between two spider coils versus the link distance $d$. We compare network analyzer measurements at $500\unit{kHz}$ in an office corridor to the signal model \eqref{eq:SignalModelGlobal} with various calibrations (which are based on fitting the measurements).}
  \label{fig:SingleLinkMeas}
\end{figure}

The factor $\xi \in \mathbb{C}$ with $|\xi| \leq 1$ represents attenuation and phase shift due to imperfect matching. Its value must be determined by calibration; in the uncalibrated case we assume $\xi = 1$ (perfect lossless matching). To appreciate this aspect it is worthwhile to look at the encountered physical reality in Fig.~\ref{fig:SingleLinkMeas}. It shows network analyzer measurements of $h$ over $d$ for a pair of coaxial coils ($\oFIX\an = \e = \oFIX\ag$). Note that $h$ is equal to the measured $S_{21}$-parameter as both coils are matched to the $50\,\Omega$ reference impedance. We first look at small distances $d < 2\unit{m}$ and observe that the uncalibrated model has an offset in magnitude and phase (the $90^\circ$ phase shift stems from the law of induction). This is compensated by calibration of $\xi$, yielding a model that accurately fits the measurements for small $d$.

For larger distances, about $d \geq 3$, the measured $h$ levels off and does not follow a simple path-loss law anymore. It seems to approach a limit value instead. We suspect this is due to multipath propagation of the radiated longwaves which are reflected and scattered at buildings, mountains, and the ground. The resultant magnetic field can be assumed spatially constant across our entire setup because it is well-established \cite{Molisch2012} that the spatial variations of such a scatter field are on the order of a wavelength (here $\lambda \approx 600\unit{m}$). This motivates the introduction of the constant field vector $\fieldNLOS{} \in \mathbb{C}^3$ (per anchor) to account for multipath propagation and the limit value $h \approx \coeff \fieldNLOS{}\Tr \oFIX\ag$ for larger $d$. We do not attempt a geometry-based calculation; instead $\fieldNLOS{}$ shall be \textit{determined by calibration} (in the uncalibrated case we assume $\fieldNLOS{} = {\bf 0}$). As seen in Fig.~\ref{fig:SingleLinkMeas}, a calibration of both $\xi$ and $\fieldNLOS{}$ leads to a great fit between measurement and model at all distances of interest.\footnotemark{}


\footnotetext{We can't say with certainty whether our approach predominantly captures radiative multipath propagation or also other effects in $\fieldNLOS{}$, although the following observations support the former. The limit value magnitude increases rapidly with $f_\mathrm{c}$ as expected for a radiation effect in this regime. Near-field distortions as cause would show strong path loss according to the image model and thus contradict the observed limit value. Still a near-field effect could play a role, e.g., the building's reinforcement bar mesh acting as passive relays. For localization, however, this aspect is secondary: the model is beneficial as it empirically improves the fit between measurement and model.}

\section{Localization Algorithm}\label{sec:algo}
With a total of $N$ anchors we want to estimate $\p\ag$ and $\oFIX\ag$ given the measurements ${\bf h}\meas = [\,h_1\meas \ldots h_N\meas]\Tr$. Thereby, $h_n\meas \in \mathbb{C}$ is the channel coefficient between the agent and the $n$-th anchor with $n = 1, \ldots, N$. All anchor-dependent quantities ($\p\an$, $\oFIX\an$, $\xi$, $\fieldLOS{}$, $d$, $\e$, $\bnf$, \ldots) are now subject to index $n$. The signal model \eqref{eq:SignalModelGlobal} describes the physical reality ${\bf h}\meas$ up to a model error $\ModelErrorVec \in \mathbb{C}^N$, i.e.
\begin{align}
{\bf h}\meas = {\bf h}\model + \ModelErrorVec .
\label{eq:VectorSignalModel}
\end{align}
We consider ${\bf h}\model = [\,h_1\model \ldots h_N\model]\Tr$ as a function of $\p\ag$ and $\oFIX\ag$ while the model parameters $\coeff$, $\fieldNLOS{}$, $\p\an$, $\oFIX\an$ are fixed. Consequently, least-squares estimates of $\p\ag$ and $\oFIX\ag$ can be obtained by minimizing 
$\|{\bf h}\meas - {\bf h}\model \|^2$  subject to $\|\oFIX\ag\|^2 = 1$.
%
A spherical parametrization of $\oFIX\ag$ yields an unconstrained 5D non-linear least squares problem which can be approached with a gradient-based solver. This is however computationally expensive and often misses the global optimum \cite{DumphartPIMRC2017}. For vast improvements in these regards we instead use the weighted least squares (WLS) algorithm proposed in \cite{DumphartPIMRC2017}, of which an adaption to our complex model \eqref{eq:SignalModelGlobal} is summarized in the following. The Levenberg-Marquardt algorithm with some initialization $\p\ag$ is applied to the 3D non-linear least squares problem
\begin{align}
\hat\p\ag = \argmin_{\p\ag} \| {\bf W} ({\bf h}\meas - {\bf A}\Tr  \hat\oFIX\ag) \|^2
\label{eq:WLS3DpStep}
\end{align}
where $\bf W$, $\bf A$ and $\hat\oFIX\ag$ are all functions of $\p\ag$. 
In particular,
${\bf W} = \diag(d_1^3 / \alpha_1 \ldots d_N^3 / \alpha_N)$ is an adaptive weighting that balances the contributions of the different anchors (which can vary by orders of magnitude) and ${\bf A} \in \mathbb{C}^{3 \times N}$ is composed of the columns $\coeff_n ( \fieldLOS{,n} + \fieldNLOS{,n} )$.
The orientation estimate 
$\hat\oFIX\ag\! = \argmin_{\oFIX\ag} \!\| {\bf W}\! ({\bf h}\meas - {\bf A}\Tr \oFIX\ag) \|^2$ subject to $\|\oFIX\ag\|^2 = 1$
is computed after any update of $\p\ag$. This constrained linear least squares problem has an efficient solution based on an eigenvalue decomposition. For the details we refer to \cite{DumphartPIMRC2017,Gander1980}. It shall be noted that we study location estimation from an instantaneous observation ${\bf h}\meas$ without temporal filtering.

Robust location estimation via error minimization requires a signal model that accurately describes ${\bf h}\meas$ well for any true $\p\ag$ and $\oFIX\ag$. This in turn calls for a calibration. 
\section{System \& Coil Design}\label{sec:setup}
Our localization approach measures ${\bf h}\meas \in \mathbb{C}^N$, i.e. all agent-to-anchor channel coefficients. For the presented study this is implemented by connecting all nine matched coils (the agent and eight anchors) to a multiport network analyzer (Rohde \& Schwarz ZNBT8) via coaxial cables and measuring the respective S-parameters. This way we establish phase synchronization, which would be a challenge (and potential source of error) if done wirelessly. The network analyzer is configured to use a $6\unit{dBm}$ probing signal and $5\unit{kHz}$ measurement bandwidth.

\newcommand\myHeight{4.65cm}
\newcommand\myHeightset{4.65cm}
\begin{figure}[!ht]
  \centering
  \subfloat[Setup with 8 anchors, agent on positioner, NWA]{
    \includegraphics[height=\myHeight,trim=4 0 3 2,clip=true]{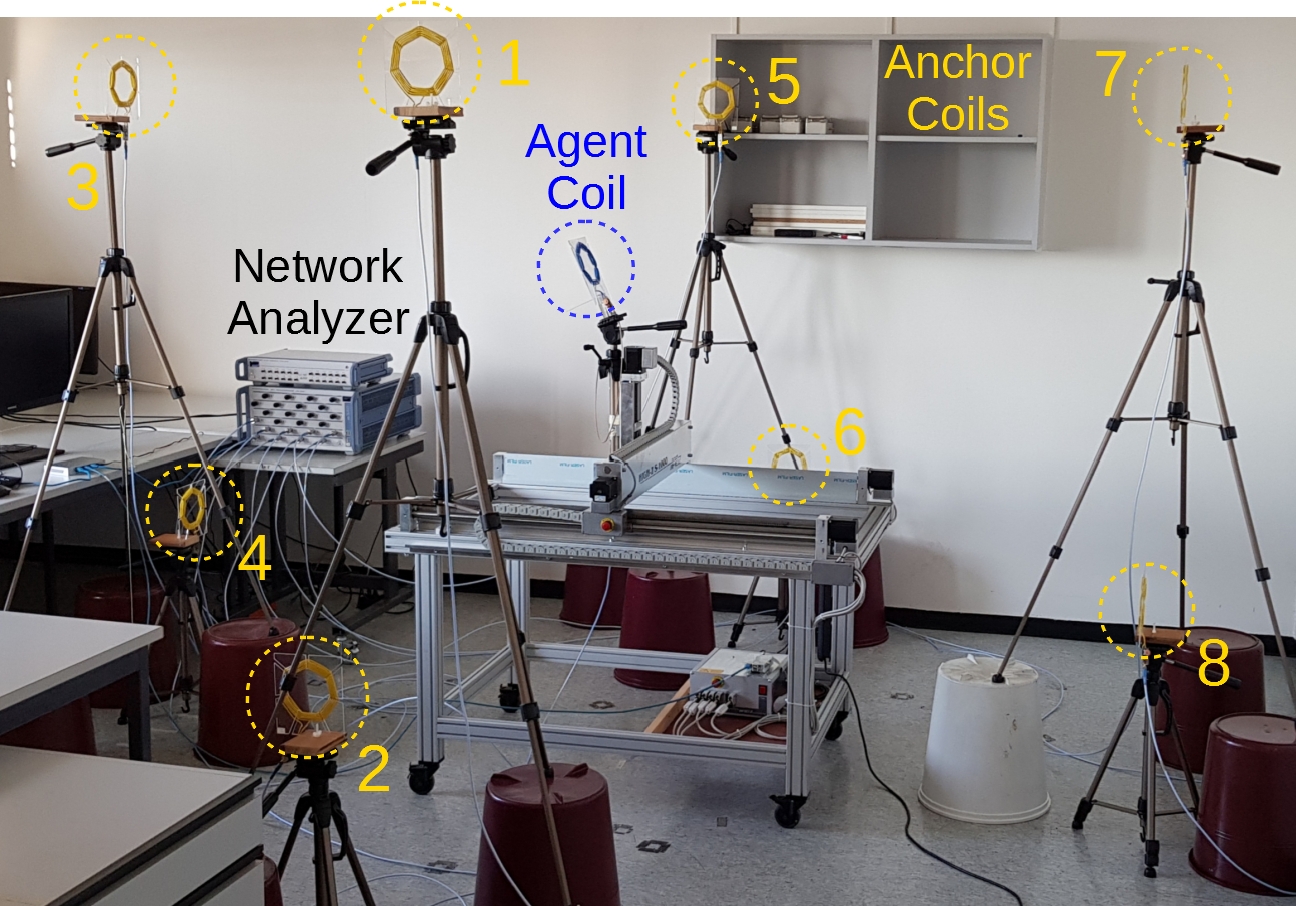}
    \label{fig:PhotoSetup}\!\!}
  \subfloat[An anchor coil]{
    \includegraphics[height=\myHeight,trim=0 0 0 0,clip=true]{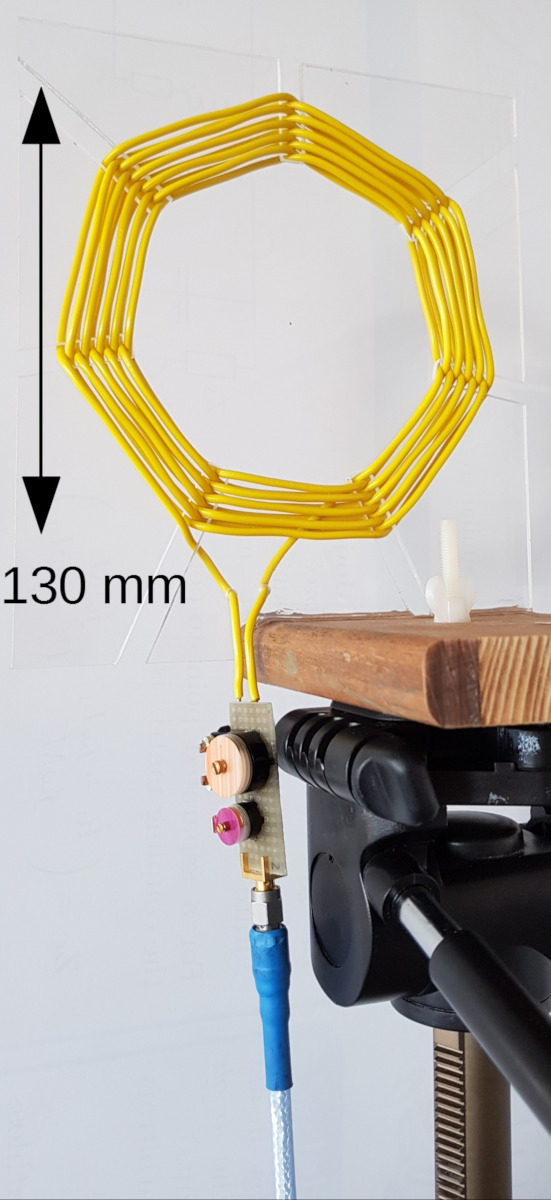}
    \label{fig:PhotoCoil}}\\[-3mm]
       \subfloat[Setup sketch, side view]{\!
      \includegraphics[height=\myHeightset,trim=0 0 0 0,clip=false]{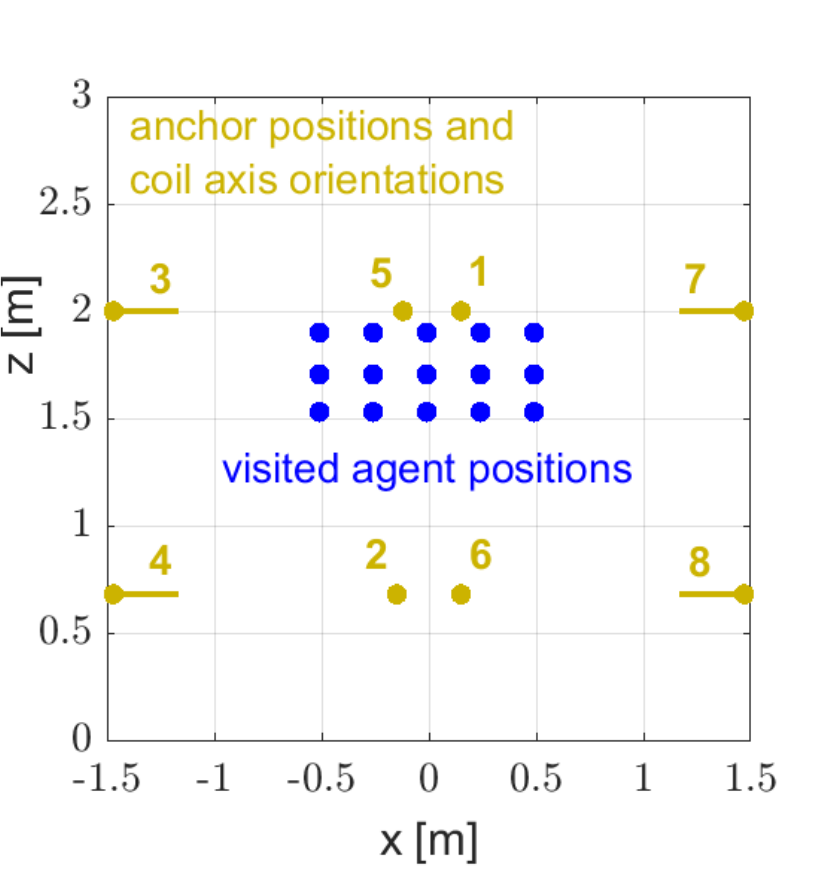}
    \label{fig:sideview}\!\!}
     \subfloat[Setup sketch, top view]{
      \includegraphics[height=\myHeightset,trim=0 0 0 0,clip=false]{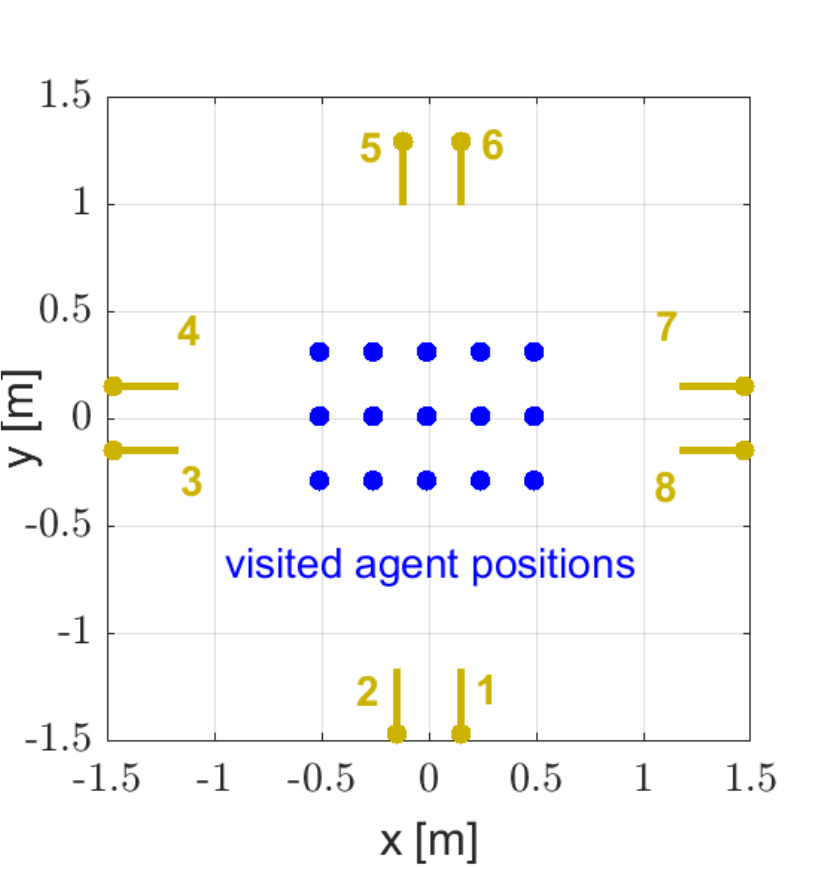}
    \label{fig:topview}}
  \caption{Setup with $N=8$ stationary anchor coils (yellow) and a mobile agent coil (blue). The anchor-to-agent measurements are acquired with the multiport network analyzer shown in (a). The anchors are distributed on the border of a $3\unit{m}\times3\unit{m}$ area as shown in (a), (c) and (d).
  Note the alternating elevation of the anchors: the anchors $n \in \{1,3,5,7\}$ are $2\unit{m}$ above the floor, those with $n \in \{2,4,6,8\}$ are at $68\unit{cm}$. This is to establish ample anchor spread in all three dimensions. In (c) and (d), the $45$ predefined agent positions $\p\ag$ which are visited via the positioner device are shown in blue. Not shown are the six different agent orientations $\oFIX\ag$ that will be assumed.}
  \label{fig:Setup}
\end{figure}

At each node, we use a spiderweb coil wounded on a plexiglas body with 10 turns each, i.e. $\nu\ag = \nu\an = 10$. The inner and outer coil diameters are $100\unit{mm}$ and $130\unit{mm}$, respectively. We use fairly thick wire with $1\unit{mm}$ diameter to keep ohmic resistance low. The operation frequency is $f = 500\unit{kHz}$, which is well below the coil self-resonance frequency of about $13\unit{MHz}$. The choice of $500\unit{kHz}$ results from the trade-off between having large channel gains at smaller distances and suppressing radiation.%
\footnote{This was concluded by comparison to a $1\unit{MHz}$ design. When doubling $f_\mathrm{c}$ in this regime, the decaying graph at smaller $d$ in Fig.~\ref{fig:SingleLinkMeas} gains $3\unit{dB}$ ($+6\unit{dB}$ from doubled induced voltages and $-3\unit{dB}$ from the skin effect increasing the coil resistance) but for the leveled graph at larger $d$ we suspect a $15\unit{dB}$ gain (it seems to scale like the far-field term in \eqref{eq:DirectPath}). This effectively reduces the reach of the decaying  graph which however holds particularly valuable location information. For an experiment at the $13.56\unit{MHz}$ ISM frequency, radiation was dominant and localization was impossible.
}
Since the wire is orders of magnitude shorter than the $600\unit{m}$ wavelength, it is safe to assume a spatially constant current distribution which is necessary for signal model \eqref{eq:SignalModelGlobal}
 to hold. The coils have $0.36\,\Omega$ resistance and $17.3\,\mu\mathrm{H}$ self-inductance (measured).

At each coil we use a two-port network to match the coil impedance to the connecting $50\,\Omega$ coaxial cable. In particular, each matching network is an L-structure of two capacitors with high Q-factor (otherwise the added resistance would pose a performance bottleneck). We use conjugate matching, which maximizes $|h_n|^2$ and resonates the coil.

The agent coil is mounted on a positioner device (HIGH-Z S-1000 three-axis positioner) which is controlled via Matlab. This allows for accurate and automated adjustment of $\p\ag$. In particular, the agent coil is mounted to the positioner with a \mbox{2-DoF} joint. This allows to adjust orientation $\oFIX\ag$, whereby the resultant change of coil center position $\p\ag$ is duly considered.

The acquisition time for a channel vector ${\bf h}\meas$ is about $14\unit{ms}$ and the execution time of the localization algorithm (WLS, single initialization) is approximately $31\unit{ms}$ on an Intel Core i7-7500U processor, allowing for up to 22 estimates per second or even 32 when acquisition and computation are parallelized. This amounts to great real-time capabilities.



%
\section{Observed Results}\label{sec:eval}
In this section we evaluate and study the localization accuracy of the presented system after a thorough calibration process. The positioner device was placed in the middle of the anchors (see Fig. \ref{fig:Setup}) and was used to visit $45$ different positions $\p\ag$ of the mounted agent coil, as illustrated in Fig. \ref{fig:sideview} and \ref{fig:topview}. These positions were visited six times with six different orientations $\oFIX\ag$. In particular, we chose for $\oFIX\ag$ the canonical x, y and z directions as well as directions at $45^\circ$ to the axes in the xy, yz and xz planes. In total, we establish $45 \cdot 6 = 270$ different agent deployments $(\p\ag,\oFIX\ag)$. At each of these $270$ deployments the S-Parameters were measured with the network analyzer, resulting in a set of channel vector measurements $\mathbf{h}^{\mathrm{meas}}_i$ for $i=1, \hdots, 270$. In order to avoid overfitting we partition the measurements into an evaluation subset $i \in \{1,3,5,\ldots\}$ and a calibration subset $i \in \{2,4,6,\ldots\}$. 

We consider an \textit{essential calibration} which adjusts $\xi$ and $\fieldNLOS{}$ for all anchors and a \textit{full calibration} which additionally tunes the anchor positions and orientations to compensate minor inaccuracies of the installation. Hence, essential and full calibration adjust $8$ and $13$ real-valued parameters, respectively, per anchor.
All $\xi$ and $\fieldNLOS{}$ parameters are calibrated with least-squares estimation per anchor. Afterwards, if a full calibration is conducted, each individual anchor position and orientation is calibrated individually by maximum a posteriori (MAP) estimation with informative priors.

\begin{figure}[!htb]
\includegraphics[trim={0 -5 0 0},clip=false,width=\columnwidth]{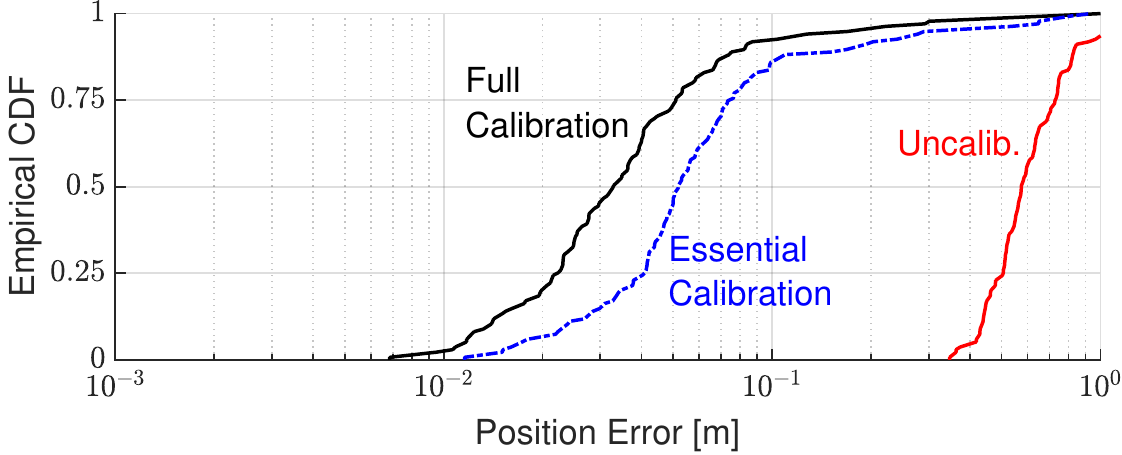}
\includegraphics[trim={0  8 0 0},clip=false,width=\columnwidth]{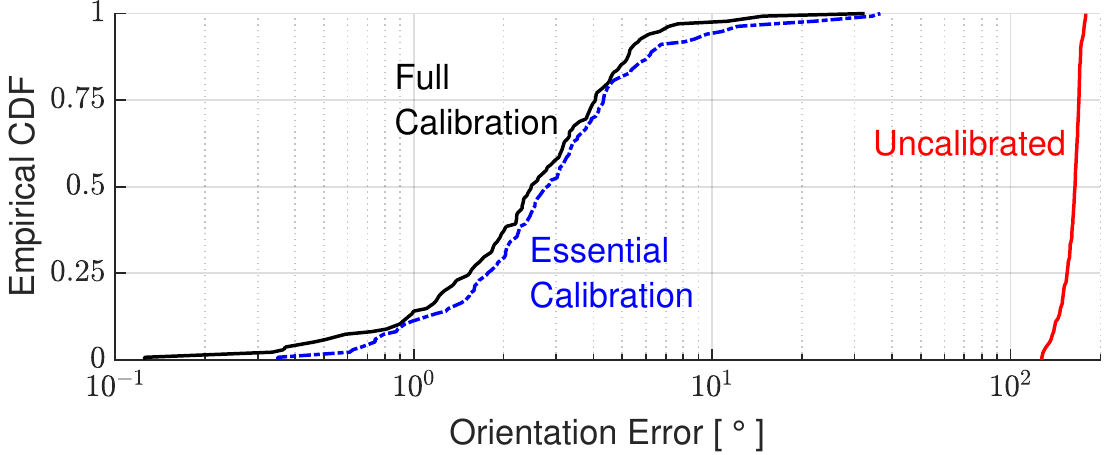} 
\caption{Performance of the 3D localization system in terms of position and orientation estimation error. The results are shown in terms of cumulative distribution function (CDF) of the localization error over the different visited agent deployments $(\p\ag,\oFIX\ag)$ (the $135$ deployments of the evaluation set) and for different calibrations.}
\label{fig:EvalCDF}
\end{figure}

The localization system then applies the WLS algorithm \eqref{eq:WLS3DpStep} with $3$ random initializations to the  evaluation subset of the channel vector measurements.  Fig. \ref{fig:EvalCDF} shows the resulting localization error statistics in terms of cumulative distribution functions (CDFs).
We observe that a median position error of about $5.2 \unit{cm}$ is obtained with the essential calibration, whereas full calibration even yields a $3.2 \unit{cm}$ median error and a $90  \% $ confidence to be below $8.3 \unit{cm}$. The median error of orientation estimation is below $3^{\circ}$ for both calibrations. It can be observed that the uncalibrated system often estimates the opposite direction because of an assumed reverse polarity, resulting in errors close to $180^{\circ}$.
We did not obtain accuracy improvements from a whitening operation on the measurements based on the empirical model error covariance matrix.

\renewcommand\myHeightset{5.7cm}
\begin{figure}[!ht]
  \centering
       \subfloat[Model error (anchor 1)]{
      \includegraphics[height=\myHeightset,trim=0 0 0 0,clip=true]{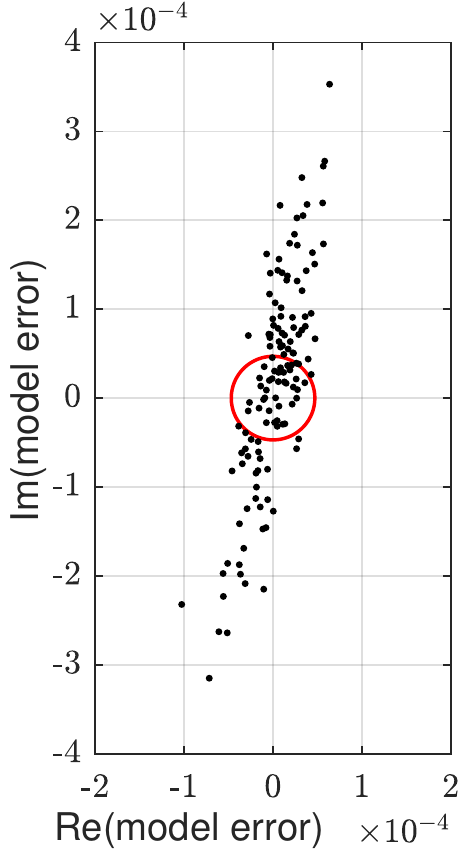}
    \label{fig:ScatterPlotModel}\!}
     \subfloat[Measurement error for stationary coils]{
      \includegraphics[height=\myHeightset,trim=0 0 0 0,clip=true]{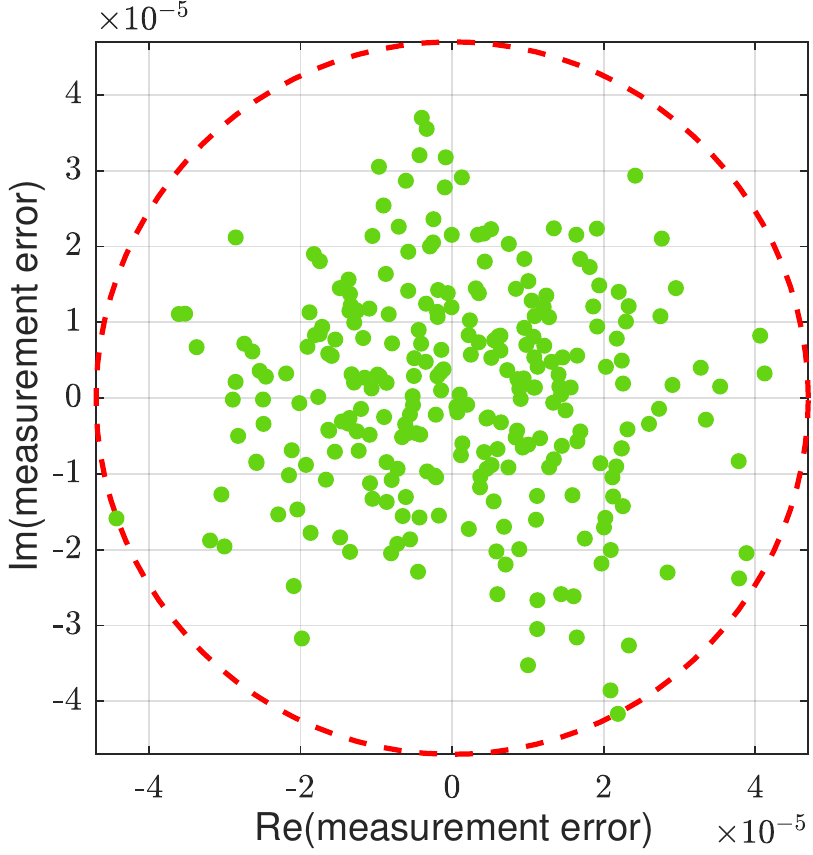}
    \label{fig:ScatterPlotMeas}}
\caption{Scatter plot (a) shows the first component (anchor $n=1$) of the model error $\ModelErrorVec_i = {\bf h}\meas_i - {\bf h}\model_i$ of the fully calibrated system. Thereby, $i$ indexes the visited agent deployments (evaluation subset $i = 1,3,5,\ldots,269$). The points are close to a line at about $80^\circ$ to the real axis, an angle that can also be observed in Fig.~\ref{fig:SingleLinkMeas} at small distances: in the near-field the phase is always close to $\pm 90^\circ$ depending on polarity, with a small offset due to mismatch $\xi$. Now $\xi$ multiplies ${\bf h}\meas_i$ and ${\bf h}\model_i$ and thus also $\ModelErrorVec_i$, resulting in the observed angle of the cloud.
Scatter plot (b) shows the errors of channel coefficient measurements acquired over time between stationary coils whereby the mean was subtracted. This shows the error magnitude due to noise, interference, and quantization. Observe how the standard deviation of this error carries over to the model error in (a) and how it spreads the associated cloud (red circle). Please note the different axis scales and that (a) and (b) were obtained with completely different methodology. Thus there is no one-to-one correspondence between data points of the two plots.}
\label{fig:ScatterPlot}
\end{figure}

While the achieved accuracy may be sufficient for various applications, one might expect better accuracy from a thoroughly calibrated system. The system seems to face a similar bottleneck as the related work and we want to investigate the cause. As outlined in Sec.~\ref{sec:algo}, the key to accurate localization with this approach is a small model error $\ModelErrorVec = {\bf h}\meas - {\bf h}\model$. Therefore, the observed residue model error is a key quantity for the study of accuracy limits as it reflects unconsidered effects. Finding its dominant contribution corresponds to isolating the system's performance bottleneck and might allow to even further improve the accuracy. Fig. \ref{fig:ScatterPlotModel} shows and discusses the realizations of the model error $\ModelErrorVec_i$ after full calibration for the different agent deployments $i$ (for brevity we depict only the first component, i.e. $n=1$).
We analyze the relative model error $|\ModelError_{n,i}|\, \big/ \,|h\meas_{n,i} | = |h\meas_{n,i} - h\model_{n,i} |\, \big/ \,|h\meas_{n,i} |$ over the entire evaluation set and find a median relative error of $0.055$ and 90th percentile of $0.302$.

As first possible cause we investigate measurement noise. Its statistics can be observed in the fluctuations of channel gain measurements about their empirical mean for a stationary agent as illustrated in Fig.~\ref{fig:ScatterPlotMeas}. By comparing the deviation magnitudes in Fig.~\ref{fig:ScatterPlotModel} and Fig.~\ref{fig:ScatterPlotMeas}, it can be seen that measurement noise is not the limiting factor for our system and that the system is not SINR limited. This also means that noise averaging of the measurement error would not improve the accuracy significantly whereas it would harm the real-time capabilities of the system.

Another possible performance bottleneck are the errors due to the assumptions that underlie the signal model of Sec.~\ref{sec:model}. Even in free space, the employed model is exact only between two dipoles or between a thin-wire single-turn circular loop antenna and an infinitesimally small coil. The actual coil apertures and spiderweb geometry are however neglected by the model. We evaluate the associated performance impact with the following procedure. Instead of acquiring ${\bf h}\meas_i$ by measurement, we synthesize it with a simulation that would be exact between thin-wire spiderweb coils in free space. In particular, we solve the double line integral \cite[Eq.~11]{DumphartWCNC2019} numerically based on a 3D model of the coil geometry and the feed wire seen in Fig.~\ref{fig:PhotoCoil}. Subsequently, we apply the same calibration and evaluation routines to the synthesized ${\bf h}\meas_i$ as previously. This results in a relative model error with median $0.0114$ and a 90th percentile of $0.0615$, which are significantly lower than the practically observed values. Hence, this aspect does not pose the performance bottleneck.

As third possible cause investigate the model error due to unconsidered nearby conductors which react with the generated magnetic field. To this effect, we tested the impact that the ferroconcrete building structures of the setup room have on single link measurements. Indeed, moving the coils closer to a wall or the floor can affect ${\bf h}\meas$ considerably. In an experiment of two coplanar coils (parallel to the floor, i.e. vertical orientation vectors) at $d = 2\unit{m}$ link distance, we compared the impact of different elevations by first choosing $0.5\unit{m}$ and then $1\unit{m}$ elevation above the floor for both coils. Although the links should be equal according to the free space model, we observed a relative deviation of $0.11$ between the two $h\meas$. This exceeds the $0.055$ median relative model error of our fully-calibrated system, which is a strong hint that nearby ferro-concrete building structures pose a significant performance bottleneck for our localization system.

\section{CRLB-Based Accuracy Projections}\label{sec:bounds}
For the sake of the following study, we assume a Gaussian distribution of the model error $\ModelErrorVec$. The adequacy of this assumption is supported by Fig.~\ref{fig:ScatterPlot} (however we do not assume circular symmetry as it would contradict Fig.~\ref{fig:ScatterPlotModel}). Consequently, we consider the probabilistic model
\begin{align}
\mtx{l}{\re\ \ModelErrorVec \\ \im\ \ModelErrorVec} \sim \mathcal{N}({\bf 0}, \boldsymbol\Sigma )
\label{eq:Gaussian}
\end{align}
where $\boldsymbol\Sigma \in \mathbb{R}^{2N \times 2N}$ is the covariance matrix of the real-valued stack vector. It can be found either empirically or with theoretical models. In this section, we compare the achievable performance for different technically relevant cases of $\boldsymbol\Sigma$.

As a tool for this comparison, we use the position error bound (PEB), which is the Cram\'er-Rao lower bound on the root-mean-square position error. The PEB is an established tool in radio localization \cite{ShenTIT2010Pt1} and handily allows for projections of estimation accuracy. For the error statistics \eqref{eq:Gaussian} with known $\boldsymbol\Sigma$, the PEB for an agent deployment $\p\ag, \oFIX\ag$ is computed as \cite[Eq.~3.31]{Kay1993}
\begin{align}
\mathrm{PEB}(\,\p\ag,\oFIX\ag) = \sqrt{\mathrm{trace}\big\{ \big(\FIM_{\estParam}^{-1}\big)_{1:3,1:3} \big\}}
\label{eq:PEB}
\end{align}
with the $5 \times 5$ Fisher information matrix
\begin{align}
\FIM_\estParam &= \fp{{\bf g}\Tr}{\estParam} \,\boldsymbol\Sigma^{-1} \Big(\, \fp{{\bf g}\Tr}{\estParam} \Big)\Tr , &
{\bf g} &= \mtx{l}{ \re\,{\bf h}\model \! \\ \im\,{\bf h}\model \! } .
\label{eq:FIM}
\end{align}
The 5D estimation parameter $\estParam = [\p\ag\Tr,\,\phi,\,\theta]\Tr$ describes the agent deployment; $\phi$ and $\theta$ are the azimuth and polar angles. The Jacobians comprise derivatives of \eqref{eq:SignalModelGlobal} that follow from an extension of \cite[Appendix B]{DumphartPIMRC2017} and are omitted for brevity.

We determine the value of $\boldsymbol\Sigma$ as follows for the described cases of interest.
\begin{enumerate}
\itemsep0em 
\item Empirically from model errors $\ModelErrorVec_i$ after full calibration.
\label{li:Actual}
\item Empirically from the measurement fluctuations observed while the agent is stationary.
\label{li:HypMeas}
\item Same as \ref{li:HypMeas}) but the transmitting agent coil was disconnected and replaced by a $50\,\Omega$ termination. 
\label{li:HypTerm}
\item Same as \ref{li:Actual}) but the ${\bf h}\meas_i$ were obtained by free-space EM simulation for thin-wire spiderweb coils instead of actual measurements, cf. Sec.~\ref{sec:eval}.
\label{li:HypSynth}
\item Thermal noise and typical background noise \cite[Fig.~2]{RadioNoiseITU2016} picked up by the anchors at $f_\mathrm{c} = 500\unit{kHz}$. The assumed spatial correlation model uses a Bessel function of $kd$ times the inner product of the two associated anchor orientations (same as in \cite{DumphartWCNC2019}).
\label{li:HypThermalBack}
\item Independent thermal noise of power $N_0 B$ at each anchor. We assume the minimum noise spectral density at room temperature, i.e. $N_0 = -174\unit{dBm}$ per $\unit{Hz}$. The bandwidth is $B = 5\unit{kHz}$ as specified in Sec.~\ref{sec:setup}.
\label{li:HypThermal}
\end{enumerate}

\begin{figure}[!ht]
\includegraphics[trim={0 0 0 0},clip,width=\linewidth]{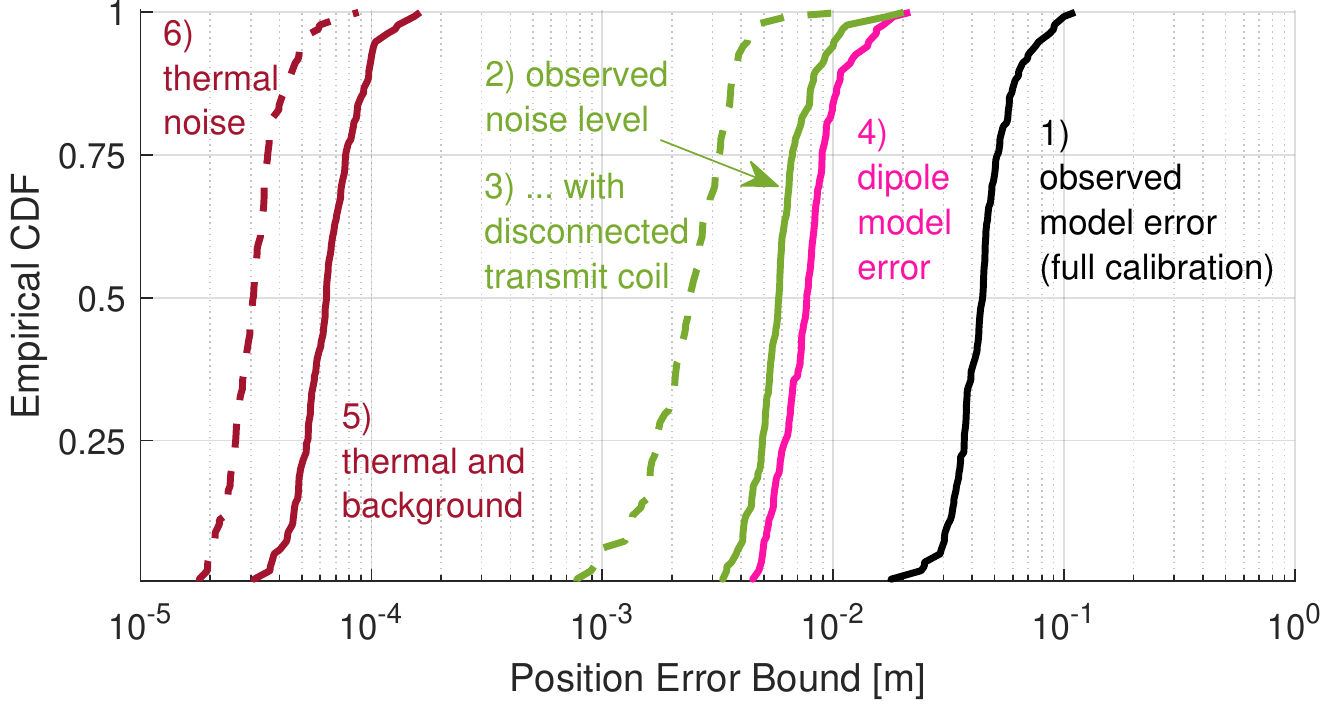}
\caption{Distribution of the position error bound (PEB) over various agent deployments within the $3\unit{m} \times 3\unit{m}$ room, evaluated for six different hypotheses on the statistics of the model error $\ModelErrorVec = {\bf h}\meas - {\bf h}\model$. The results serve as accuracy projections for various (idealistic) circumstances. We suspect that the significant differences between cases \ref{li:HypMeas} and \ref{li:HypTerm} are caused by the way the network analyzer adapts the probing signal to the link, e.g. adding dither noise to mitigate ADC non-linearity and quantization error.}
\label{fig:PebCdf}
\end{figure}

We evaluate $\mathrm{PEB}(\,\p\ag,\oFIX\ag)$ for the same agent deployments as in Sec.~\ref{sec:eval} and for the different described values of $\boldsymbol\Sigma$. The results are shown in Fig.~\ref{fig:PebCdf}.
First of all, the PEB for case~\ref{li:Actual} matches the practically achieved accuracy in Fig.~\ref{fig:EvalCDF} well (however it shall be noted that the PEB applies to the RMSE and not to single error realizations). The results for the cases \ref{li:HypMeas} and \ref{li:HypTerm} indicate the performance limit assuming that noise, interference and/or quantization determine the achievable accuracy. This case would allow for sub-$\mathrm{cm}$ accuracy. The PEB-results for case \ref{li:HypSynth} also exhibit sub-$\mathrm{cm}$ accuracy in most cases but are slightly worse than cases \ref{li:HypMeas} and \ref{li:HypTerm}. This case \ref{li:HypSynth} is particularly important as it represents the accuracy limit of parametric location estimation based on an analytical signal model such as \eqref{eq:SignalModelGlobal}. This indicates that sub-$\mathrm{cm}$ accuracy is infeasible for this approach and a system of our scale, even in a distortion- and interference-free environment. The cases \ref{li:HypThermalBack} and \ref{li:HypThermal} use noise power estimates from communication theory and yield projections between $20$ and $200\um$. These are vastly optimistic as they would require an extremely high-resolution ADC and a precise and well-calibrated signal model that accounts for any appreciable physical detail.

\balance
\section{Summary \& Conclusions}\label{sec:summary}
We presented a magneto-inductive system with 8 flat anchor coils to localize an arbitrarily oriented agent coil. After thorough calibration we achieve an accuracy better than $10\unit{cm}$ in $92\%$ of cases in an office environment. 
The underlying signal model was complemented by the radiative direct path and a means to account for multipath propagation.
We quantitatively investigated the potential accuracy-limiting factors and identified distortions due to conductive building structures to be dominant. We project that $1\unit{cm}$ accuracy is possible in a distortion-free environment or by accurately modeling any appreciable impact of nearby conductors (which would however vastly complicate the system deployment). Much better than $1\unit{cm}$ accuracy seems to be infeasible for magneto-inductive indoor localization via parameter estimation based on an analytical signal model.

\section*{Acknowledgment}
We would like to thank Manisha De for supporting the system development.

\IEEEtriggeratref{0}
\bibliographystyle{IEEEtran}
\bibliography{GD}

\end{document}